\documentclass{aa}
\usepackage{amsmath}
\usepackage{graphicx,txfonts,natbib,color}
\bibpunct{(}{)}{;}{a}{}{,}

\begin{document}

\title{Shifting of the resonance location for planets embedded in circumstellar disks}
\author{F. Marzari\inst{1}}
\institute{
  Dipartimento di Fisica, University of Padova, Via Marzolo 8,
  35131 Padova, Italy
}
\titlerunning{Resonance shifting}
\authorrunning{F. Marzari}
\abstract 
{In the early evolution of a planetary system, a pair of planets may be captured 
in a mean motion resonance while still embedded in their nesting circumstellar disk. 
}
{ The goal is to estimate the direction and amount of shift in the semimajor axis of the 
resonance location due to the disk gravity as a function of the gas density and mass of the planets. The stability of the resonance lock when the disk dissipates is also tested. 
}
{The orbital evolution of a large number of systems is numerically integrated within 
a three-body problem in which the disk potential is computed as a series of expansion. 
This is a good approximation, at least over a limited amount of
time. 
}
{Two different resonances are studied: the 2:1 and the 3:2. In both cases the 
shift is inwards, even if by a different amount, when the planets are massive 
and carve a gap in the disk.  For super--Earths, the shift is instead outwards.
Different disk 
densities, $\Sigma,$ are considered and the resonance shift depends almost linearly 
on $\Sigma$. The gas dissipation leads to destabilization of a significant 
number of resonant systems, in particular if it is fast. 
}
{The presence of a massive circumstellar disk may significantly affect the 
resonant behavior of a pair of planets by shifting the resonant location and
by decreasing the size of the stability region. The disk dissipation may 
explain some systems found close to a resonance but not locked in it. 
}


\maketitle

\section{Introduction} 
\label{intro}

During the early phases of evolution of a planetary system it may happen 
that two planets are trapped in a mean-motion resonance.  
By inspecting known multi-planet systems, a 
significant number of planets can be found  very close to a commensurability and, 
in particular, the closeness to low-order resonances such as 2:1 and 3:2 is
frequently observed  \citep{wright2011,fab2012}. 
For example, the 3:2 resonance is possibly present in HD 45364 \citep{rein2010}, KOI 55 (Kepler 70)
\citep{charpinet2011}, and many other Kepler systems \citep{steffen2013}.
Well-known systems with planets 
suspected to be in the 2:1 resonance are  Gliese
876 and HD 82943. 

The assembly of a resonant configuration may be the
result of formation or dynamical evolution of the system.
Multi-planet systems may directly form in a chain of resonances like
Kepler-223   (Kepler-223's two innermost planets are in a 4:3 resonance, 
the second and third are in a 3:2 resonance and, finally,  
the third and fourth are in a 4:3 resonance) leading to 
a tightly packed orbital configuration \citep{mills2016}.
Direct in situ formation in a dense disk \citep{hansen2013} or inside-out  formation  \citep{tan2014}  
may lead to resonant pairs even if the lock 
may be disrupted later on during the evolution of the system.

A mechanism often invoked to explain resonant configurations is 
convergent migration where two planets are moved inward
together by tidal interaction with the circumstellar disk
(e.g., \cite{masset2001}, \cite{lee2002},
\cite{adams2005},  \cite{thommes2005}, \cite{beauge2006},  
\cite{beauge2006}, \cite{crida2008}).
In a typical planet migration scenario, a giant planet drifting
inwards is reached from outside by a less massive
planet migrating at a faster pace. Depending on the disk 
density, the couple can be trapped in different mean motion resonances 
\citep{angelo2012}, in particular the 2:1 or the 3:2. This may lead 
to a period of coupled inward/outward migration until the disk 
dissipates. The `grand tack' scenario \citep{walsh2011} is a possible 
application of this mechanism. It is also possible that the inner giant 
planet triggers the formation of a tightly packed planetary system 
with all the members in resonance. 

Once trapped in resonance, the common gap determines the 
coupled migration rate \citep{angelo2012}. However, even a different dynamical effect 
comes into play when the planets are in resonance. In addition to the apsidal 
precession rate induced by the resonant perturbations, there is a significant 
contribution forced by the overall gravitational force of the gas disk. 
This can significantly affect the dynamics of the planets in resonance by shifting the
location of potential commensurabilities between them (Tamayo et al. 2015) and, 
during the disk dissipation, may also lead to resonance escape.

Various analytical approaches have been developed to compute this precession rate. The initial
physical assumptions at the basis of all these models are equivalent, but they gradually depart
from one another in the subsequent way of manipulating the equations and in the approximations 
adopted to derive a simple handy expression (\cite{mestel1963}, \cite{ward}, \cite{binney2008},
\cite{silsbee2015}, \cite{fonta}). However, these formalisms have been applied in most cases to compute the 
circulation rate of a single planet, while in the case of two planets in resonance, the 
situation is more complex. In the majority of resonant configurations  an apsidal 
libration is also present and the change in the secular circulation frequency due to the 
disk potential affects the 
system of the two planets as a whole, not each body individually. Therefore it is not easy 
to predict the resonance shift because it is due to the interaction between the 
secular circulation frequency at the resonance and the circulation forced by the 
disk potential on each planet. 
 For this reason a numerical 
approach is adopted here where the orbits of the planets are computed with the addition
of the gravitational force of the disk acting on each planet.  
The formalism developed by \cite{ward} is used
and the series expansions developed to compute the gravitational potential of the disk 
are numerically computed at a high order. The shift in the resonance location due to the 
disk gravity is computed for different values of the disk density and planet masses. 
The parameter space is huge, so here the exploration is limited to a few cases that provide information on the important dynamical processes related to the effects of the disk 
potential on the planet precession rate. In \cite{fonta} the comparison between the 
apsidal precession rate predicted by the analytical approach of \cite{ward} 
has been compared to that computed by hydrodynamical simulations performed with the 
code FARGO \citep{Masset} for a single planet of different masses. 
A good agreement has been observed between analytical and numerical results \citep{fonta} 
implying that the 
angular momentum exchange, related to the gas-planet interaction and 
causing planet migration, does not significantly influence the resonance location 
which is instead mostly 
related to the apsidal frequency change due to the disk gravity. 

In Section 2 the critical arguments of a planetary resonance are described, while 
in Section 3 the analytical formulas used to compute the disk potential are 
briefly summarized. Section 4 is devoted to the numerical algorithm and to the definition 
of the initial conditions. The results of the numerical modeling for the 2:1 
resonance and for different planetary masses are described in Section 5, and 
Section 6 is dedicated to the 3:2 resonance. Finally, a discussion of the results 
and the conclusions are 
given in Section 6.

\section{Planets in resonance}
\label{resonance}

The focus of this paper is on first-order mean motion resonances 
like the 2:1 and 3:2 which are the most frequently encountered 
in exoplanetary systems. From a computational point of view, 
the best way to 
determine when two planets are in resonance is to 
check the behavior of the two critical angles:

\begin{align}
\label{eq:wgap}
\sigma_1 = (j+1) \lambda_2 - j \lambda_1 - \varpi_1 ,\\
\sigma_2 = (j+1) \lambda_2 - j \lambda_1 - \varpi_2, 
\end{align}

where $\lambda_i$ is the mean longitude of planet {\it i} while 
$\varpi_i$ is its apsidal longitude. 
Outside the resonance, both $\sigma_1$ and $\sigma_2$  circulate 
from $0^o$ to $360^o$. When the planets are in resonance, 
at least one 
of the critical arguments is a libration regime and  oscillates about either $0^o$ or 
$180^o$ \citep{bea_mi_2003,callegari2004}. When both $\sigma_1$ and $\sigma_2$ 
librate, a state of apsidal libration is met where the circulation 
frequencies of the apsidal longitude of the two planets match. 

The easiest way to test if the two planets are 
in resonance during an N-body numerical simulation is to define two stripes 
around both $0^o$ and $180^o$ with a small width (in 
all simulations a half-width of $5^o$
is chosen). 
If, during the numerical integration, both the  critical angles cross 
these stripes then the system is assumed to be out of 
resonance. Until this condition is met, the system is assumed to be resonant.
This is a brute force method but it is very efficient when 
handling the outcome of a large number of numerical simulations of 
potentially resonant systems. 

From the above expressions of the critical arguments it can be 
deduced that the location of the resonance is affected by any
change in the apsidal frequencies of the 
planets due to an external perturbation such as the disk 
gravity. Since the latter alters the circulation frequency of 
$\dot{\varpi_i}$, then 
a change in the 
circulation period of $\lambda_2$ is required 
to restore the resonant lock (the 
orbit of the inner planet is fixed in all 
models, so the frequency of 
$\lambda_1$ does not change). Only a 
new frequency of $\lambda_2$, and then a different 
value of the outer planet semimajor axis, 
is able to maintain
$\sigma_i$ in a libration motion. It is noteworthy that 
the disk gravity affects the inner and outer planet to a different 
extent and the shift in the frequency of  $\lambda_2$ of the outer planet 
indirectly includes the change in the frequency of $\dot{\varpi_1}$.
This is because the planets are locked in resonance and react to the 
disk gravity as a single entity.

\section{Computation of the disk potential}
\label{potential}

In all models, a thin disk is considered with a surface gas density profile  $\Sigma (r)$ 
parameterized as a power law of the form:

\begin{equation}
\Sigma(r) = \Sigma_0 r^{-p}
.\end{equation}

The disk gravitational potential is computed using the formula of 
\cite{ward}:

\begin{align}
\label{eq:wardU}
U^W(r) = &  2\pi G r \Sigma(r)\sum_{k=0}^{+\infty}A_{k} \biggl[\frac{(4k+1)}{(2k+2-p)(2k-1+p)} \notag \\
&-\biggl(\frac{1}{2k+2-p}\biggr)\biggl(\frac{R_{in}}{r}\biggr)^{2k+2-p} \\   & -\biggl(\frac{1}{2k-1+p}\biggr)\biggl(\frac{r}{R_{t}}\biggr)^{2k-1+p}\,\biggr] \notag
\end{align}

In the above equation $r$ is the radial distance from the star, $R_{in}$ and $R_{out}$ 
are the inner and outer borders of the disk, respectively, and 
$A_{k}=[(2k)!/2^{2k}(k!)^{2}]^{2}$. 

If the planets are massive enough, they may open a common gap  and 
a more accurate approach for the computation of the disk potential is given 
by the following equation \citep{ward}:

\begin{align}
\label{eq:wgap}
U^{W} &= 2\pi G\Sigma(r)\sum_{k=0}^{+\infty}A_{k} \biggl[\frac{1}{(2k+2-p)}\biggl(\frac{D_{in}}{r}\biggr)^{2k+2-p}\biggl(1-\biggl(\frac{R_{in}}{D_{in}}\biggr)^{2k+2-p}\biggr) \notag \\
&+\frac{1}{(2k-1+p)}\biggl(\frac{r}{D_{out}}\biggr)^{2k-1+p}\biggl(1-\biggl(\frac{D_{out}}{R_{t}}\biggr)^{2k-1+p}\biggr)\biggr]. 
\end{align}

The inner and outer borders of the gap are $D_{in}$, $D_{out}$, respectively. 
To properly use this equation we need a rough estimate of the 
values of both $D_{in}$ and $D_{out}$. As an order of magnitude we adopt the equation for
the size of the gap carved by a single giant planet given in \cite{isella2016}:

\begin{equation}
W \sim C \cdot R_H 
,\end{equation}

where the constant $C$ can range from 4 to 8 and $R_H$ is the Hill's sphere. In the numerical modeling, two different values for $C$ were tested: either 3 or 4. No significant differences were observed 
in the outcome, therefore in all the simulations $C=4$ is adopted. In the numerical computation of the 
disk potential we extend the summation of the series expansion
up to $k = 100$. Some tests with a higher number of terms  show that no significant contributions 
to the potential come from higher-order terms at the cost of a considerable 
slowing down of the computations.

\section{The numerical setup}
\label{model}

The parameter space of two planets in resonance is huge, so only 
a few representative scenarios are considered. 
Two different combinations of planet masses are considered:  first a pair of massive planets 
where the inner planet has the mass of Jupiter (the inner one, $m_1 = M_J$) 
and the outer one that of Saturn 
($m_2 = M_S$).
This choice of the mass ratio  is dictated by the migration-driven resonant trapping 
mechanism which, to be efficient, requires the outer body to be less massive than the 
inner one. A second less massive case is contemplated  where  
the mass of the inner planet is set to $m_{1} = 5 M_E$  
(a super--Earth) while that of the outer one is $m_{2} = 3 M_E$.  
The initial semimajor axis of 
the inner planet $a_1$ is fixed to 5 au for the massive planets case and 
to 5 and 1 au in the super--Earths case. The semimajor axis of the second planet $a_2$
is varied within a wide range around the exact resonant value. The orbits of 
the planets are assumed to be coplanar to the disk and the eccentricity is 
randomly sampled between 0 and 0.2. The orbital angles (mean anomaly, pericenter argument and 
longitude of node) of the inner planet are set to zero while those of the outer 
planet are randomly chosen between 0 and $2 \pi$. 
To define the range within which 
to look for resonant systems, we use a rough estimate of the resonant 
width for circular orbits: 

\begin{equation}
{\Delta n \over n_R}  \simeq \mu^{2/3} 
,\end{equation}

applicable to first order resonances, where $n_R$ is the mean motion at the exact commensurability
while $\mu$ is the mass ratio.  We estimate from the above equation a value of 
$\Delta a$ and multiply it by 10 in order to also cover the models where  the disk
potential shifts inwards or outwards with respect to the exact resonant location. 

As a reference value for the gas density, the minimum mass solar nebula is adopted
\citep{chiang2010}:

\begin{equation}
\Sigma(r) = 2200 \times  r^{-3/2} \quad g/cm^2
.\end{equation}

According to \cite{chiang2013} a ‘minimum-mass extrasolar nebula’ (MMEN)  can be 
derived from the data of the Kepler mission neglecting significant migration of planets. 
Their average MMEN is approximately five times more massive than the MMSN and from their data 
even more massive disks could potentially be found around solar-type stars.
Similar results are obtained also by \cite{raymond2014}. Assuming that in any 
case  massive disks can 
be encountered, in the simulations 
two different values for $\Sigma_0$ are adopted. 
Either $\Sigma_0 = \Sigma_{MMSN} = 2200 \quad g/cm^2$ is used 
or a massive disk is modeled with $\Sigma_0 =  5 \cdot \Sigma_{MMSN}$.
The shift in resonance for intermediate values of $\Sigma_0$
can be extrapolated from these cases. The scenario explored here assumes that 
planets can form very rapidly before significant dissipation of the disk. 
They are trapped in resonance either by convergent migration or by direct
formation in a resonant configuration. Different dynamical paths leading to resonance 
can be envisaged in particular in multi-planet systems.

When modeling the dynamical evolution of 
two planets in resonance embedded in a circumstellar disk, the best approach would be to exploit a
hydrodynamical code which includes the evolution of the gas. However, this approach 
would require heavy numerical computations simply for determining the stability of a
single system. While searching for the averaged shift in the resonant location, a large
number of systems have to be integrated since not all systems that are stable 
for a given value of the disk density $\Sigma$ are stable also for a different value. Furthermore,  
it is also interesting to test the potential shrinking (or widening) of the 
resonance volume in phase space as a function of $\Sigma$. 
For these reasons a statistical approach is to be preferred where thousands of systems 
are numerically integrated to evaluate the changes in the resonant structure for 
different disks.  To pursue this objective it is not possible to 
exploit hydrodynamical simulations and an N--body approach  
where the disk potential is computed with the 
equations of \cite{ward} would allow a more accurate sampling of 
the parameter space. Here we compute a large number of different 
static resonant configurations  where 
the planets do not migrate.  The 
stability of the resonance lock is investigated over $10^5$ yrs. 
It is possible to include in the numerical models a dissipative term simulating the 
migration of the system but this would render the estimate of the shift in semimajor 
axis of the planets in resonance more difficult to evaluate. During migration, 
the secular frequencies will slightly change due to the semimajor axis variation 
and the contribution of the disk potential 
will change with time. For this reason, to evaluate the local amount of shift in the 
resonance location 
it is better to use a static configuration. In addition, as shown in \cite{fonta}, 
the comparison of the analytical formulas with numerical simulations performed with the
hydrodynamical code FARGO \citep{Masset} have shown very good agreement. It is 
then expected that also in a more complex model the apsidal precession rate is 
dominated by the disk potential and well modeled by the analytical formalism of \cite{ward}.

Once the dependence of the resonant phase space on different
values of $\Sigma$ and on different masses and semimajor axes of 
the planets has been explored, it is interesting to test whether the disk dissipation may 
destroy the resonant lock. With this goal in mind, the 
same two-planet systems have been integrated  with a model where the disk density  $\Sigma$ 
exponentially decreases with time:

\begin{equation}
\Sigma (r,t) = \Sigma_0 r^{-p} e^{-t /\tau}
,\end{equation}

where $\tau$ is the dissipation timescale. As a good compromise 
between the need for fast numerical integrations and reasonable values
for the dissipation timescale of protoplanetary disks, 
$\tau$ has been set to $5 \times 10^5$ yr. The number of stable systems 
at the end of the numerical integrations is then compared to the 
`static' case to test how often a resonant system can be destabilized 
by the change in the apsidal frequencies related to the disk dissipation. 

\section{The 2:1 resonance}
\label{res21}

The first resonance to be studied is the 2:1 which was 
analyzed for the first time by \cite{lee2002} in connection with the 
discovery of the exoplanet system GJ 876. 
This resonance will be investigated in distinct physical 
models in the forthcoming Sections to test how the semimajor axis 
separation between the two planets depends on their mass and on
the gas density. 

\subsection{Shift in the location of the resonance: massive planets}
\label{res21shift}

\begin{figure}[hpt]
  \includegraphics[width=\hsize]{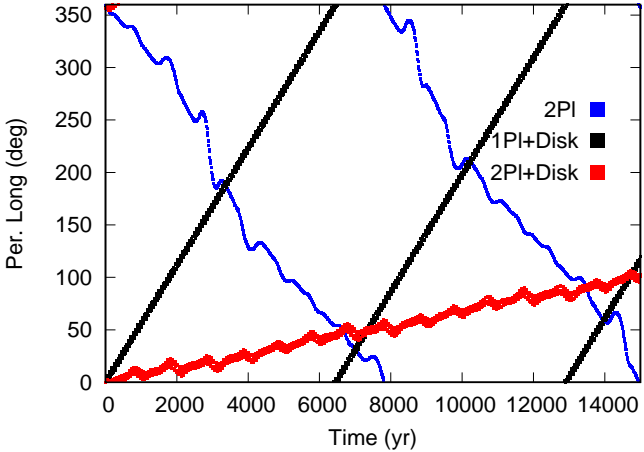}
  \caption{\label{CIRCULATION} 
Evolution with time of the pericenter longitude $\varpi$ of the 
inner more massive planet as a function of time in three different models,
all with a gap in the disk. 
The blue line shows the evolution of $\varpi$ in a pure three-body problem where the 
mutual gravitational perturbations of the planets determine the retrograde apsidal longitude circulation
of the inner one.
The black line 
illustrates the circulation of the inner planet $\varpi$ affected by the disk potential only (the second 
planet is not included) which is positive, as expected \citep{fonta}. 
The full model where both the planets and the disk potential are considered 
is shown by the red line evolving with a positive frequency because of the 
dominating effect of the disk potential.
}
\end{figure}

When two planets are placed in resonance, the apsidal longitude 
circulation frequency of the inner and
outer one, in absence of the disk potential, are determined by the  mutual gravitational 
perturbations. Since they are in resonance, these frequencies cannot be computed with the 
usual linear secular theory but they are determined by the resonant interactions. When the disk potential 
is switched on, it acts on both planets affecting the circulation of the 
apsidal longitude of each body by different amounts since the planets orbit at different locations 
with respect to the disk or, when present, to the disk gap. 
This leads to a shift in the resonance location 
which depends on the contribution of the disk potential to the 
apsidal circulation of the two bodies. In most cases, the two planets have also a 
common average rate of apsidal precession which will depend on the disk potential 
acting on both planets. In this case, the scenario appears even more complex since the 
disk potential affects the couple of bodies locked in resonance. 
In Fig.\ref{CIRCULATION} the precession of the apsidal longitude of 
the inner planet (the more massive) is shown in three different scenarios. In the first 
a pure three-body problem is numerically integrated and only the gravitational perturbations of the planets are
taken into account. In the second the disk potential is applied to both planets assuming 
they create a common gap in the disk. Finally, we integrate the orbit of the inner planet
alone 
assuming the same potential of the disk as in the previous run (including then the same gap) to 
evaluate the effect of the disk potential on a single body. 
Figure \ref{CIRCULATION} clearly shows that the negative circulation frequency of $\varpi$ 
in the pure three-body problem becomes positive 
when the gravitational perturbation of the disk is included and this forces an inward shift of the 
outer planet to maintain the resonance lock.

\begin{figure}[hpt]
  \includegraphics[width=\hsize]{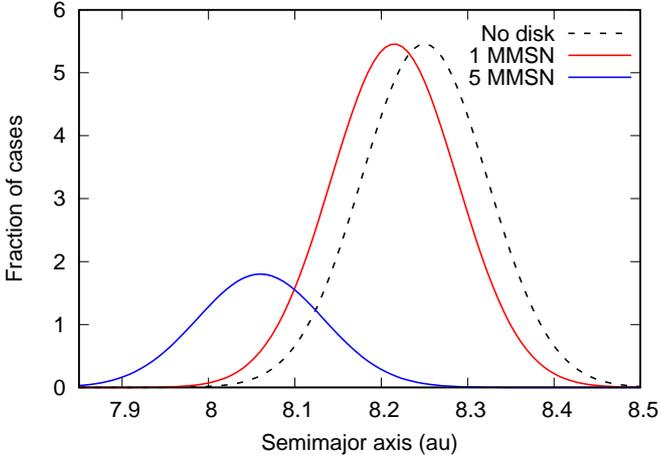}
  \caption{\label{JS_21_SHIFT_MJ}  Distribution of the initial semimajor axes of 
the outer planet of a massive pair ($m_{1} = 1 M_J \quad m_{2} = 1 M_S$)
trapped in a 2:1 resonance for different values of the disk gas density $\Sigma$. For 
higher densities the resonant region is shifted inwards. The volume of the phase 
space is also reduced and the size of the resonance region shrinks. This is 
revealed by the smaller area of the Gaussian curve fitting the histogram. 
}
\end{figure}

This behavior affects all the resonant systems, and we expect an inward shift of the 
outer planet location for all the resonant systems, which depends on the strength of the 
disk gravity. 
In Fig.\ref{JS_21_SHIFT_MJ} the distribution of the initial semimajor axis 
of the outer planet for all the sampled resonant systems is illustrated. Three different cases
are compared: a reference case where $\Sigma = 0$ (pure three-body problem),  and two models with the 
surface density set to match 1 MMSN and 5 MMSN, respectively.  
Each curve shown in Fig.\ref{JS_21_SHIFT_MJ} is the outcome of a fit to the 
histogram
of the number of cases versus the initial semimajor axis of the outer planet.
The values of the frequencies of the histogram have been normalized so that the 
underlying area is equal to 1. 
A least square fit to the histogram curve is performed with a Gaussian function:

\begin{equation}
f(a) = \frac {1} {\sigma \sqrt{ 2 \pi} } e^{- \frac {1} {2} \left (\frac {(a- \bar a)} {\sigma}\right )^2}
,\end{equation}

\begin{figure}[hpt]
  \includegraphics[width=\hsize]{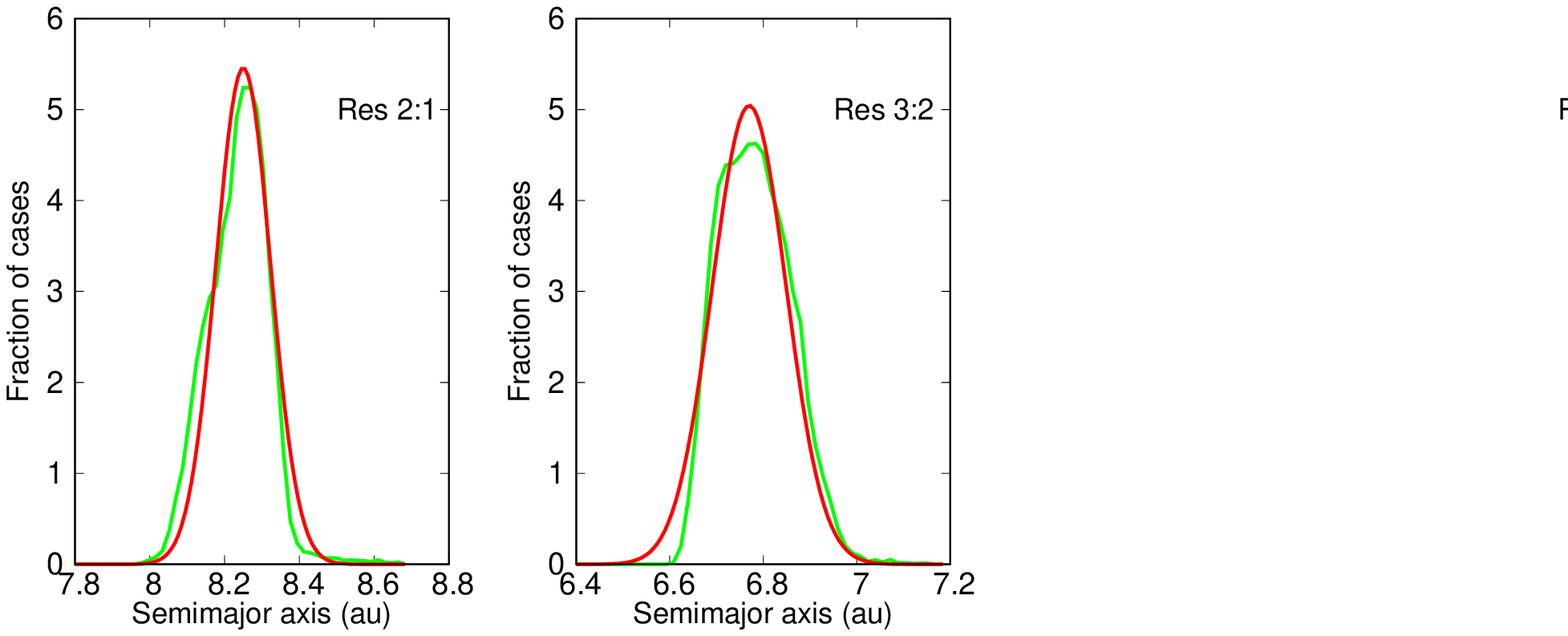}
  \caption{\label{FIT_GAUSS}  Gaussian fits to the distribution in semimajor 
axis of the outer planet in models where the potential of the disk 
is neglected (pure three-body problem). On the left panel the fit is shown for the 
2:1 resonance, while, on the right panel, the same fit is shown for the 
3:2 resonance.
}
\end{figure}

where $\bar a$ is the value of the semimajor axis at the center of the resonance and 
$\sigma $ estimates the dispersion around the center. In Fig.\ref{FIT_GAUSS} the Gaussian 
fit is shown in the pure three-body model for both the 2:1 and 3:2 resonances showing a 
very  good match between the analytical curve and the numerical density function. 

The Gaussian fit, normalized (in area)
to one, is further multiplied by a coefficient $f$  given by the ratio between the 
total number of cases when the disk potential is included, divided by 
the total number of cases in the pure three-body problem. 
The semimajor axis range of the 
outer planet, randomly explored while seeking for resonant systems, is the same in all three 
cases ($\Sigma_0 =0$, $\Sigma_0 =  1 \cdot \Sigma_{MMSN}$ and $\Sigma_0 = 5 \cdot \Sigma_{MMSN}$).
Therefore,
by counting the number of resonant systems found during the random exploration
it is possible to estimate 
the size of the resonant region and the efficiency of trapping in the 
three different cases. If we denote the total number of resonant systems found at the 
end of the random search as  $n_0$  when 
$\Sigma = 0$, the pure three-body case,  then the 
size of the resonant region for $\Sigma_0 =  1 \cdot \Sigma_{MMSN}$  
is reduced by a factor $n_1 / n_0$ where $n_1$ is the final number of
resonant cases when $\Sigma_0 =  1 \cdot \Sigma_{MMSN}$.
In the same way, the size of the resonant region is reduced by 
$n_5 / n_0$, where $n_5$ is the final number of
resonant cases, when $\Sigma = \Sigma_{5MMSN}$. 

For higher values of 
the disk surface density $\Sigma$, the center of the resonant region is shifted 
inwards by an increasing amount. This is due to the increase of the circulation 
frequency of the 
planet apsidal precession forced by the disk gravity. To  maintain the
libration of
$\sigma_1$ and/or $\sigma_2$, the outer planet approaches the inner one 
increasing its mean anomaly frequency. In most of our
models the planets are also in apsidal resonance and the frequency of the
apsidal longitude of the outer planet matches that of the inner one. As a
consequence, the secular system responds in a unitary way to the disk
perturbations.

There is also a shrinking of the volume in the phase space where stable 
resonant systems (at east over $1 \times 10^5$ yr) can be found for increasing 
$\Sigma$. This reduction is 98\% for $\Sigma = \Sigma_{1MMSN}$ and 33\% for 
$\Sigma = \Sigma_{5MMSN}$. The disk potential, when strong enough, is able to 
destabilize a significant fraction of potential resonant systems reducing the 
number of possible stable cases. The shift in the apsidal longitude circulation frequency
pushes the planets closer to each other increasing their mutual 
interaction and causing a weakening of the resonance lock. 


\subsection{Resonance break up due to disk dissipation}
\label{res21dissi}

\begin{figure}[hpt]
  \includegraphics[width=\hsize]{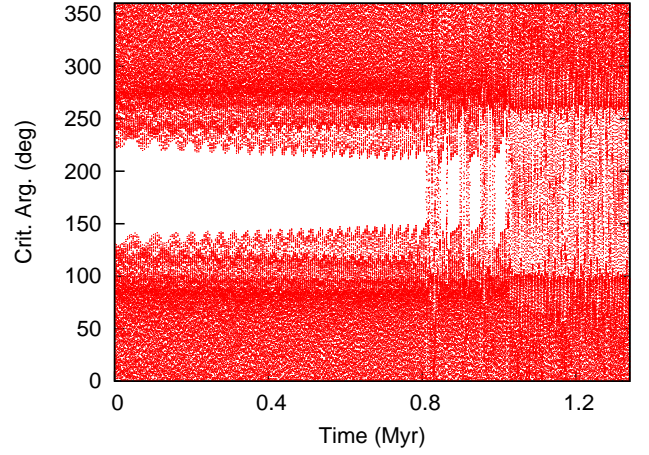}
  \caption{\label{JS_21_DISSI}  Time evolution of the 
$\sigma_1$ critical argument of the 2:1 resonance during the 
dissipation of the disk. The libration slowly turns into 
circulation (for both $\sigma_1$ and $\sigma_2$) as the 
density of the gas declines leading to the break up of the 
resonance.  The initial value of $\Sigma$ is equal to 
$\Sigma_{1MMSN}$. 
}
\end{figure}

During the gas dissipation due to photo-evaporation, viscous accretion, and 
stellar winds, the resonance lock can be broken. In the models where we 
adopt an exponentially declining $\Sigma$, some systems break out of 
the resonance and both critical arguments circulate from then on. An example is shown 
in Fig.\ref{JS_21_DISSI} where the resonance breakup occurs after about 0.8 Myr
with an exponential dissipation coefficient $\tau = 5 \times 10^5$ yr. 

By performing a statistical exploration, through integration of the same systems with and 
without dissipation, we can estimate the fraction of systems potentially trapped 
in resonance that are out of it once the disk dissipates. About 27 \% of the 
system initially in resonance is found in a circulating state after a timescale 
equal to $ 2 \tau,$ confirming that the disk dissipation may break up the 
resonance leaving the planets close to it. When the density of the disk is increased to 
$\Sigma = \Sigma_{5MMSN}$ the fraction of cases surviving the disk dissipation 
drops to 50\%. If we neglect all other effects related to the planet-disk interaction 
and we focus on the disk potential and its effects on the apsidal precession, we 
find that for higher disk densities the dissipation prevents the evolution from being 
adiabatic and the system is more easily destabilized. 

\subsection{Shift in the location of the resonance: super-Earths}
\label{res21shift}

\begin{figure}[hpt]
  \includegraphics[width=\hsize]{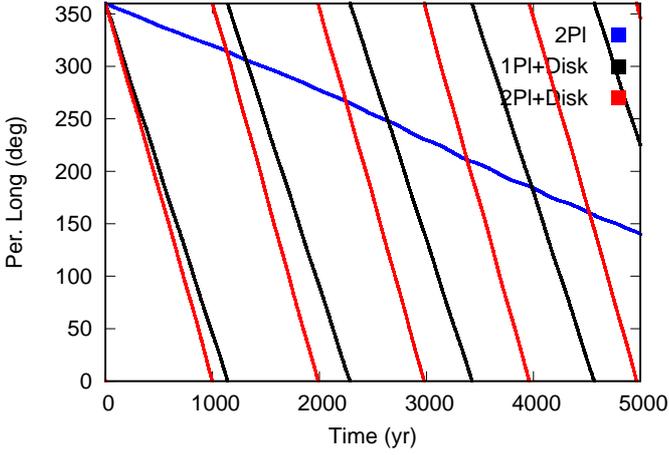}
  \caption{\label{CIRCULATION_ME}
Evolution with time of the pericenter longitude $\varpi$ in three different models as 
in Fig.\ref{CIRCULATION}. In this case the planets have five and three Earth masses, respectively,
and the inner one has a semimajor axis of 1 au. The disk potential causes a negative precession 
which adds up to that due to the mutual gravitational perturbations of the planets. As a 
consequence, the circulation rate of $\varpi$ is negative in the full model with two 
planets and the disk potential (red line), contrary to what happens for 
massive planets within a disk gap.  It is also faster compared to that observed in a pure three-body 
problem (blue line) and for a single planet (black line).  
}
\end{figure}

Here we consider two planets whose masses are five and three times the Earth mass, respectively, 
with the more massive planet in the inner orbit. They are 
supposed to be trapped in resonance during their type-I migration or formation 
process, like inside-out growth. Their masses are not 
large enough to open a gap in the disk and for this reason we adopt for the disk 
potential Eq. \ref{eq:wardU} with both planets embedded in the disk. 
The precession rate of the apsidal longitude due to the 
disk potential on a single planet is negative (black line in Fig. \ref{CIRCULATION_ME})
and concordant with the circulation induced by the resonant interaction (blue line). 
As a consequence, 
the circulation rate in the full model including the two planets and the disk 
(red line in Fig. \ref{CIRCULATION_ME})
is faster compared to that of a single planet at 1 au and to that forced 
by the resonance. This behavior is opposite to that of two massive planets 
carving a common gap in the disk where the circulation caused by the disk 
potential is opposite and dominant with respect to the negative one driven by the resonance. 
Therefore, it is expected that the shifting in the resonance location for super-Earths
is outwards rather than inwards as in the case of massive planets. 

\begin{figure}[hpt]
  \includegraphics[width=\hsize]{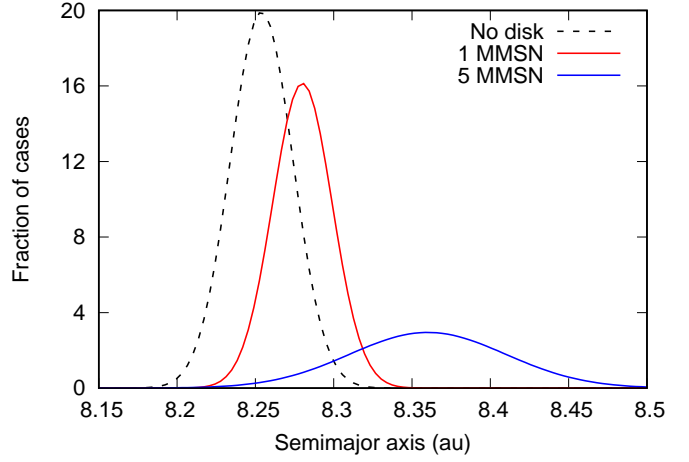}
  \includegraphics[width=\hsize]{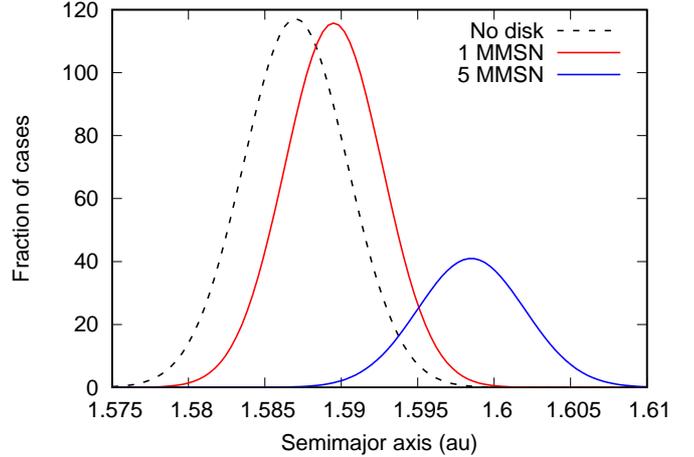}
  \caption{\label{JS_21_SHIFT_ME}  Distribution of the initial semimajor axes of 
the outer planet in a system composed of two 
super--Earths  in a 2:1 resonance for different values of the disk gas density. For 
higher densities the resonant region is shifted inwards. The upper panel 
models a pair of planets with the inner one at 5 au, while the lower panel models 
the case where the inner planet is at 1 au.  
}
\end{figure}

This is indeed observed in Fig. \ref{JS_21_SHIFT_ME} where we show the distribution of the resonant cases when the 
semimajor axis of the inner planet is 1 and 5 au, respectively. The shift is outwards in both cases,
as expected, 
but there is a significantly lower reduction of the resonance efficiency when 
$\Sigma = \Sigma_{5MMSN}$ for $a_{1} = 1$ au. The possible reason for this 
different behavior is that 
the contribution to the total gravitational potential of the inner part of the 
disk is more destabilizing for large densities. As a consequence, since in the 
`close' case the amount of gas mass in between the inner planet and the star 
is small and significantly less compared to the `far' case, the resonance is 
more stable even in the case of high gas density. 

\section{The 3:2 resonance}
\label{res32}

\begin{figure}[hpt]
  \includegraphics[width=\hsize]{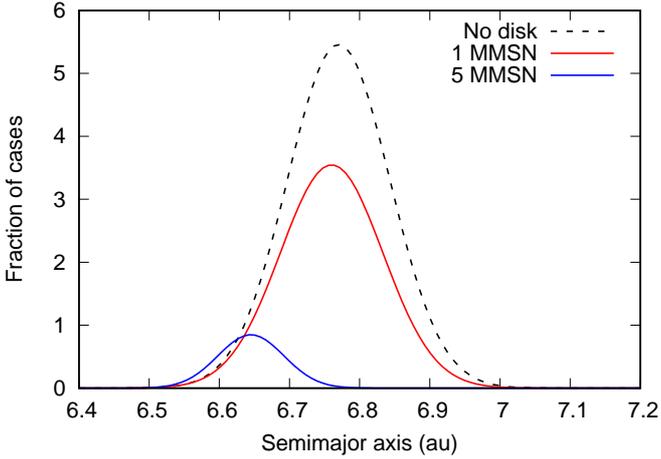}
  \caption{\label{JS_32_SHIFT}  Distribution of the initial semimajor axes of 
massive planets in the 3:2 resonance for different values of the disk gas density. 
}
\end{figure}

For Jupiter-size planets, the shift in the resonance location appears less marked for the 3:2 resonance, 
while, on the other side,  the shrink in the resonant region appears more consistent. This is in 
agreement with the hypothesis that the forced proximity of the planets induced 
by the shift in the apsidal circulation decreases the stability of the resonant lock. 
The number of 
resonant configurations is decreased by about 90\% in the case where the disk density is set to
5 MMSN in comparison to the pure N-body case (Fig.\ref{JS_32_SHIFT}).  The 3:2 resonance appears more 
sensitive to the dissipation of the disk gas and about 40\% of the systems are destabilized and 
escape from resonance before the end of the simulation for $\Sigma = \Sigma_{1MMSN}$. 
The intrinsic weakness of the 3:2 resonance to external perturbations, compared to the 2:1, 
suggests that less systems should be found trapped in it but rather are found close to it. 

\begin{figure}[hpt]
  \includegraphics[width=\hsize]{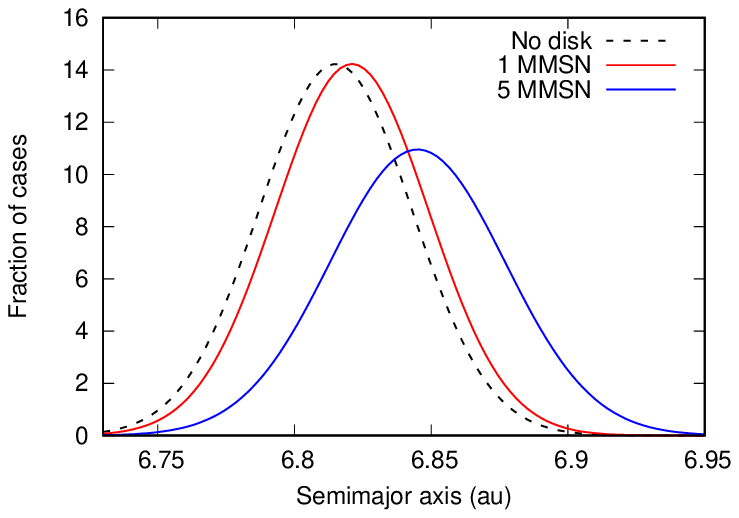}
  \includegraphics[width=\hsize]{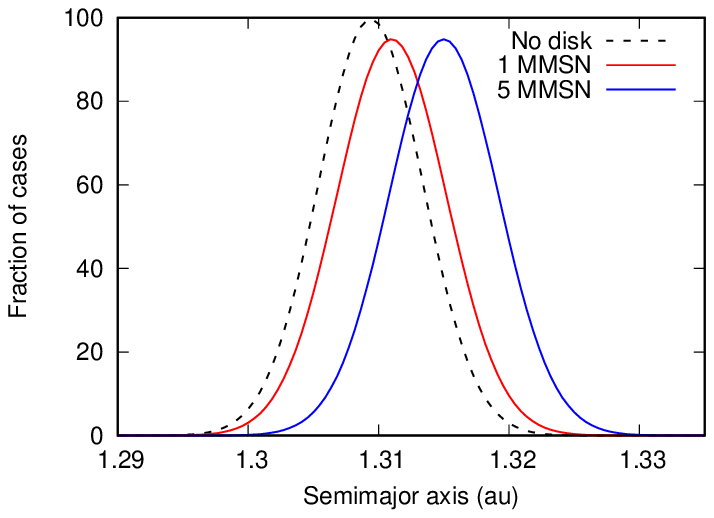}
  \caption{\label{JS_32_SHIFT_ME}  Distribution of the initial semimajor axes of
the outer planet in a system composed of two
super-Earths in the 3:2 resonance for different values of the disk gas density. 
The upper panel
models a pair of planets with the inner one at 5 au, while the lower panel models
the case where the inner planet is at 1 au.
}
\end{figure}

When  super--Earths are considered (Fig. \ref{JS_32_SHIFT_ME}),  the shift in the resonant location 
is present in both the far ($a_1 = 5$ au) and close (($a_1 = 1$ au) configurations, but the 
resonance efficiency is not significantly decreased in either case. There is a more marked decrease 
in the `far' case, possibly for the same  reasons already outlined at the end of 
Section \ref{res21shift} related to the reduced contribution of the inner part of the 
disk that may be responsible for most of the instability at higher densities. However, there is 
potentially a dependence on the mass of the planets since the instability for super-Earth pairs
is significantly less compared to giant planet pairs. This may be related to the direction of 
the shift that is outwards for massive planets and inwards for super-Earths. 

\begin{figure}[hpt]
  \includegraphics[width=\hsize]{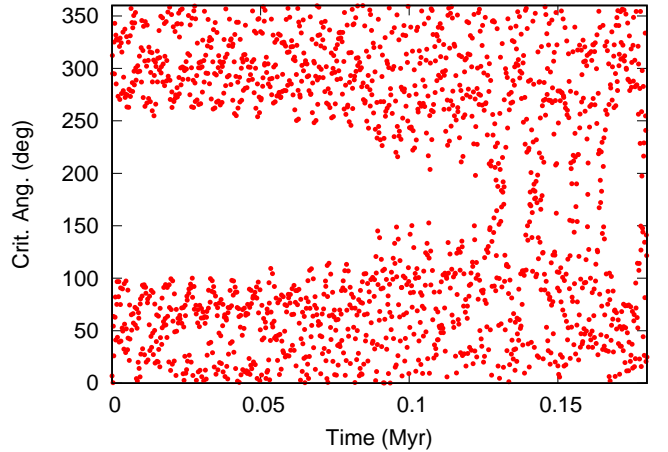}
  \caption{\label{JS_32_DISSI}  Time evolution of the 
$\sigma_1$ critical argument of the 3:2 resonance during the 
dissipation of the disk for the super-Earth case. 
The libration width grows with time and finally it turns into 
circulation (for both $\sigma_1$ and $\sigma_2$) when
the density of the gas declines. This leads to the break up of the 
resonance and a period of chaotic evolution characterized by mutual 
close encounters.  The initial value of $\Sigma$ is equal to 
$\Sigma_{1MMSN}$. 
}
\end{figure}

The dissipation of the disk affects also the stability of super-Earths and in the 
model with a declining gas density the number of cases which are destabilized is 
about 25 \% when $\Sigma = \Sigma_{1MMSN}$. In Fig. \ref{JS_32_DISSI} the behavior of the critical argument of
the resonance is shown for two super-Earths when $a_1 = 1$ au.
As in Fig. \ref{JS_21_DISSI}, 
the critical argument 
progressively increases its libration 
amplitude until circulation takes place. In some cases, immediately after the resonance 
breakup, the planets, before the end of the simulation, have mutual close encounters and 
undergo a period of chaotic 
behavior. 

\section{Discussion and Conclusions}
\label{discussion}

The coexistence of planets and the gaseous disk suggests interesting dynamical scenarios 
where the gravitational forces between the planets are perturbed by the disk 
gravity. Exchange of angular momentum may occur due to the planet perturbations 
on the disk leading to different types of migration depending on the 
mass of the planets. An additional effect is the change in the secular frequencies of 
a planet due to the disk gravitational force. This last effect is of no relevance if 
the interest is focused on the migration speed of the planets, however it may be 
of importance when dealing with two planets in resonance. The shift in the 
secular frequencies leads to a change in the resonance location of the outer planet 
once the semimajor axis of the inner one is fixed. This shift in the resonance 
location may be inwards or outwards depending on the mass of the planets and the 
formation (or not) of a common gap. 

In this paper the amount of the resonance shift is numerically estimated by a massive 
exploration of the phase space around the 2:1 and 3:2 resonances. The shift depends on 
the disk density and on the mass of 
the planets. The focus here is on Jupiter-Saturn-like couples  and 
super-Earths. When the disk is very massive (5 MMSN) the shift is consistent and 
the stability of the resonance is jeopardized by the strong attraction of the disk. 
The percentage of systems surviving in resonance decreases significantly as a function of the 
disk density for both resonances. 
This effect is less relevant for super-Earths, in particular when a `close' configuration is 
considered where the inner planet is located at 1 au from the star. In this case the perturbative
effects of the inner regions of the disk are reduced and a higher stability is granted. 

It is reasonable to expect that during the evolution of the system the disk dissipates due 
to different mechanisms, such as, photo-evaporation and viscous evolution, among others. In this 
scenario, the resonance lock may be broken and the planets are left either in  a stable 
configuration close to the resonance or they evolve chaotically during mutual close encounters 
until a planet is ejected from the system or a collision occurs. This may explain why a
significant fraction of planets are found close to a resonance but not exactly in it. 

\section*{Acknowledgments}
I thank A. Fontana and an anonymous referee for their useful comments and suggestions.

\bibliographystyle{aa}
\bibliography{biblio}

\begin{thebibliography}{28}
\expandafter\ifx\csname natexlab\endcsname\relax\def\natexlab#1{#1}\fi

\bibitem[{{Beaug{\'e}} \& {Michtchenko}(2003)}]{bea_mi_2003}
{Beaug{\'e}}, C. \& {Michtchenko}, T.~A. 2003, \mnras, 341, 760

\bibitem[{{Beaug{\'e}} {et~al.}(2006){Beaug{\'e}}, {Michtchenko}, \&
  {Ferraz-Mello}}]{beauge2006}
{Beaug{\'e}}, C., {Michtchenko}, T.~A., \& {Ferraz-Mello}, S. 2006, \mnras,
  365, 1160

\bibitem[{{Binney} \& {Tremaine}(2008)}]{binney2008}
{Binney}, J. \& {Tremaine}, S. 2008, {Galactic Dynamics: Second Edition}
  (Princeton University Press)

\bibitem[{{Callegari} {et~al.}(2004){Callegari}, {Michtchenko}, \&
  {Ferraz-Mello}}]{callegari2004}
{Callegari}, Jr., N., {Michtchenko}, T.~A., \& {Ferraz-Mello}, S. 2004,
  Celestial Mechanics and Dynamical Astronomy, 89, 201

\bibitem[{{Charpinet} {et~al.}(2011){Charpinet}, {Fontaine}, {Brassard},
  {Green}, {Van Grootel}, {Randall}, {Silvotti}, {Baran}, {{\O}stensen},
  {Kawaler}, \& {Telting}}]{charpinet2011}
{Charpinet}, S., {Fontaine}, G., {Brassard}, P., {et~al.} 2011, \nat, 480, 496

\bibitem[{{Chatterjee} \& {Tan}(2014)}]{tan2014}
{Chatterjee}, S. \& {Tan}, J.~C. 2014, \apj, 780, 53

\bibitem[{{Chiang} \& {Laughlin}(2013)}]{chiang2013}
{Chiang}, E. \& {Laughlin}, G. 2013, \mnras, 431, 3444

\bibitem[{{Chiang} \& {Youdin}(2010)}]{chiang2010}
{Chiang}, E. \& {Youdin}, A.~N. 2010, Annual Review of Earth and Planetary
  Sciences, 38, 493

\bibitem[{{Crida} {et~al.}(2008){Crida}, {S{\'a}ndor}, \& {Kley}}]{crida2008}
{Crida}, A., {S{\'a}ndor}, Z., \& {Kley}, W. 2008, \aap, 483, 325

\bibitem[{{D'Angelo} \& {Marzari}(2012)}]{angelo2012}
{D'Angelo}, G. \& {Marzari}, F. 2012, \apj, 757, 50

\bibitem[{{Fabrycky} \& {Kepler Science Team}(2012)}]{fab2012}
{Fabrycky}, D.~C. \& {Kepler Science Team}. 2012, in AAS/Division of Dynamical
  Astronomy Meeting, Vol.~43, AAS/Division of Dynamical Astronomy Meeting, 1.03

\bibitem[{{Fontana} \& {Marzari}(2016)}]{fonta}
{Fontana}, A. \& {Marzari}, F. 2016, \aap, 589, A133

\bibitem[{{Hansen} \& {Murray}(2013)}]{hansen2013}
{Hansen}, B.~M.~S. \& {Murray}, N. 2013, \apj, 775, 53

\bibitem[{Isella {et~al.}(2016)Isella, Guidi, Testi, Liu, Li, Li, Weaver,
  Boehler, Carperter, De~Gregorio-Monsalvo, Manara, Natta, P\'erez, Ricci,
  Sargent, Tazzari, \& Turner}]{isella2016}
Isella, A., Guidi, G., Testi, L., {et~al.} 2016, Phys. Rev. Lett., 117, 251101

\bibitem[{{Lee} \& {Peale}(2002)}]{lee2002}
{Lee}, M.~H. \& {Peale}, S.~J. 2002, \apj, 567, 596

\bibitem[{{Masset}(2000)}]{Masset}
{Masset}, F. 2000, \aaps, 141, 165

\bibitem[{{Masset} \& {Snellgrove}(2001)}]{masset2001}
{Masset}, F. \& {Snellgrove}, M. 2001, \mnras, 320, L55

\bibitem[{{Mestel}(1963)}]{mestel1963}
{Mestel}, L. 1963, \mnras, 126, 553

\bibitem[{{Mills} {et~al.}(2016){Mills}, {Fabrycky}, {Migaszewski}, {Ford},
  {Petigura}, \& {Isaacson}}]{mills2016}
{Mills}, S.~M., {Fabrycky}, D.~C., {Migaszewski}, C., {et~al.} 2016, \nat, 533,
  509

\bibitem[{{Moorhead} \& {Adams}(2005)}]{adams2005}
{Moorhead}, A.~V. \& {Adams}, F.~C. 2005, \icarus, 178, 517

\bibitem[{{Raymond} \& {Cossou}(2014)}]{raymond2014}
{Raymond}, S.~N. \& {Cossou}, C. 2014, \mnras, 440, L11

\bibitem[{{Rein} {et~al.}(2010){Rein}, {Papaloizou}, \& {Kley}}]{rein2010}
{Rein}, H., {Papaloizou}, J.~C.~B., \& {Kley}, W. 2010, \aap, 510, A4

\bibitem[{{Silsbee} \& {Rafikov}(2015)}]{silsbee2015}
{Silsbee}, K. \& {Rafikov}, R.~R. 2015, \apj, 798, 71

\bibitem[{{Steffen} {et~al.}(2013){Steffen}, {Fabrycky}, {Agol}, {Ford},
  {Morehead}, {Cochran}, {Lissauer}, {Adams}, {Borucki}, {Bryson}, {Caldwell},
  {Dupree}, {Jenkins}, {Robertson}, {Rowe}, {Seader}, {Thompson}, \&
  {Twicken}}]{steffen2013}
{Steffen}, J.~H., {Fabrycky}, D.~C., {Agol}, E., {et~al.} 2013, \mnras, 428,
  1077

\bibitem[{{Thommes}(2005)}]{thommes2005}
{Thommes}, E.~W. 2005, \apj, 626, 1033

\bibitem[{{Walsh} {et~al.}(2011){Walsh}, {Morbidelli}, {Raymond}, {O'Brien}, \&
  {Mandell}}]{walsh2011}
{Walsh}, K.~J., {Morbidelli}, A., {Raymond}, S.~N., {O'Brien}, D.~P., \&
  {Mandell}, A.~M. 2011, \nat, 475, 206

\bibitem[{{Ward}(1981)}]{ward}
{Ward}, W.~R. 1981, \icarus, 47, 234

\bibitem[{{Wright} {et~al.}(2011){Wright}, {Veras}, {Ford}, {Johnson}, {Marcy},
  {Howard}, {Isaacson}, {Fischer}, {Spronck}, {Anderson}, \&
  {Valenti}}]{wright2011}
{Wright}, J.~T., {Veras}, D., {Ford}, E.~B., {et~al.} 2011, \apj, 730, 93

\end{thebibliography}

\label{lastpage}

\end{document}